\documentclass{article}%
\usepackage{amsfonts}
\usepackage{amsmath}
\usepackage{amssymb}
\usepackage{graphicx}%
\providecommand{\U}[1]{\protect\rule{.1in}{.1in}}
\providecommand{\U}[1]{\protect\rule{.1in}{.1in}}
\newtheorem{theorem}{Theorem}

\newtheorem{example}[theorem]{Example}

\newtheorem{proposition}[theorem]{Proposition}
\newtheorem{remark}[theorem]{Remark}

\newenvironment{proof}[1][Proof]{\noindent\textbf{#1.} }{\ \rule{0.5em}{0.5em}}
\begin{document}

\title{Construction of coupled Harry Dym hierarchy and its solutions from St\"{a}ckel systems.}
\author{Krzysztof Marciniak\\Department of Science and Technology \\Campus Norrk\"{o}ping, Link\"{o}ping University\\601-74 Norrk\"{o}ping, Sweden\\krzma@itn.liu.se
\and Maciej B\l aszak\\Faculty of Physics, A. Mickiewicz University\\Umultowska 85, 61-614 Pozna\'{n}, Poland\\blaszakm@amu.edu.pl}
\date{March 15, 2010}
\maketitle

\begin{abstract}
In this paper we show how to construct the coupled (multicomponent) Harry Dym
(cHD) hierarchy from classical St\"{a}ckel separable systems. Both nonlocal
and purely differential parts of hierarchies are obtained. We also construct
various classes of solutions of cHD hierarchy from solutions of corresponding
St\"{a}ckel systems.

\end{abstract}


\section{Introduction}

Various relations between finite- and infinite-dimensional nonlinear
integrable systems have been investigated since the middle of 70:s in a long
sequence of papers starting from the paper \cite{bognov}, through papers
\cite{antforstef}-\cite{cKdV} (see for example \cite{mb1} for more detailed
bibliography) and many others. In all these efforts, however, the main idea
was to pass from infinite- to finite-dimensional integrable systems. This
paper is a third paper in our series of papers showing that also an opposite
way is possible: that of passing from ordinary differential equations
integrable in the sense of Arnold-Liouville to infinite-dimensional integrable
systems (soliton hierarchies). In paper \cite{bensol} we demonstrated a way of
generating commuting evolutionary flows \ from corresponding family of
St\"{a}ckel systems (that is classical finite dimensional Hamiltonian systems
quadratic in momenta and separable in the sense of Hamilton-Jacobi theory). We
presented our idea in the setting of coupled (multicomponent) KdV hierarchies
(for definition and properties of these hierarchies, see for example
\cite{AF}). In paper \cite{makrzkdv} we systematized and developed this idea
by showing how solutions of these St\"{a}ckel systems can be used for
generating various classes of solutions of cKdV hierarchies. Although both
papers have been written for the case of cKdV, similar constructions are
possible for other hierarchies as well. In this paper we demonstrate a way of
generating the coupled (i.e. multicomponent) Harry Dym (cHD) hierarchy (see
\cite{AF2},\cite{AF3}) and various classes of its solutions from a class of
St\"{a}ckel systems of Benenti type. Our method leads both to the nonlocal cHD
hierarchy as well as to purely differential cHD hierarchy, that is to a
multicomponent generalization of HD hierarchy discussed in \cite{brunelli}
(see also \cite{hunter}). The nonlocal part of cHD hierarchy has not been
discussed in \cite{AF2} at all. We also clarify and simplify some of the
results given in \cite{bensol},\cite{makrzkdv}.

The paper is organized as follows. In Section 2 we briefly remind some basic
fact about St\"{a}ckel separable systems and discuss how they are related to
corresponding Killing systems (dispersionless nonlinear PDE's of evolutionary
type defined by Killing tensors of St\"{a}ckel systems). Sections 3 and 4 are
devoted to description of nonlocal multicomponent Harry Dym hierarchy and its
various solutions, respectively. Sections 5 and 6 are devoted to local (purely
differential) cHD hierarchy.

\section{St\"{a}ckel systems and their dispersionless counterpart}

St\"{a}ckel separable systems can be most conveniently obtained from an
appropriate class of separation relations. Generally speaking, $n$ equations
of the form%

\begin{equation}
\varphi_{i}(\lambda_{i},\mu_{i},a_{1},\ldots,a_{n})=0\text{, \ }%
i=1,\ldots,n\text{, \ \ \ }a_{i}\in\mathbf{R} \label{SRalg}%
\end{equation}
(each involving only one pair $\lambda_{i},\mu_{i}$ of canonical coordinates
on a $2n$-dimensional Poisson manifold $\mathcal{M}$) are called separation
relations \cite{Sklyanin} provided that $\det\left(  \frac{\partial\varphi
_{i}}{\partial a_{j}}\right)  \neq0$. We can then locally resolve equations
(\ref{SRalg}) with respect to $a_{i}$ obtaining
\begin{equation}
a_{i}=H_{i}(\lambda,\mu),\text{ \ \ }i=1,\ldots,n. \label{hami}%
\end{equation}
with some new functions (Hamiltonians) $H_{i}(\lambda,\mu)$ that in turn
generate $n$ canonical Hamiltonian systems on $\mathcal{M}$:%
\begin{equation}
\lambda_{t_{i}}=\frac{\partial H_{i}}{\partial\mu},\ \ \mu_{t_{i}}%
=-\frac{\partial H_{i}}{\partial\lambda},\ \ \ \ i=1,...,n. \label{2.2}%
\end{equation}
All the flows (\ref{2.2}) mutually commute since the Hamiltonians $H_{i}$
Poisson commute. Moreover, Hamilton-Jacobi equations for all the Hamiltonians
$H_{i}$ are separable in the $(\lambda,\mu)$-variables since they are
algebraically equivalent to the separation relations (\ref{SRalg}).

In this article we consider a special but important class of separation
relations, namely%
\begin{equation}%
{\displaystyle\sum\limits_{j=1}^{n}}
a_{j}\lambda_{i}^{n-j}=\lambda_{i}^{m}\mu_{i}^{2}+\frac{\varepsilon}{4}%
\lambda_{i}^{k}\text{, \ \ \ }i=1,\ldots,n \label{BenSR}%
\end{equation}
with arbitrary fixed $m,k\in\mathbf{Z,}$ $\varepsilon=\pm1$ (the constant
$\frac{1}{4}$ is not essential for the construction and is only introduced for
a smoother identification our systems with the hierarchy in (\cite{AF2})). The
relations (\ref{BenSR}) are linear in the coefficients $a_{i}$ so that they
can be (globally) solved by Cramer formulas, which yields%

\begin{equation}
a_{i}=\mu^{T}K_{i}G^{(m)}\mu+\frac{\varepsilon}{4}V_{i}^{(k)}\equiv
H_{i}^{n,m,k}\text{, \ }i=1,\ldots,n\text{, \ \ }m,k\in\mathbf{Z} \label{HamB}%
\end{equation}
where we denote $\lambda=(\lambda_{1},\ldots,\lambda_{n})^{T}$ and $\mu
=(\mu_{1},\ldots,\mu_{n})^{T}$. Functions $H_{i}$ defined as the right hand
sides of (\ref{HamB}) depend on $m$ and $k$ and can be interpreted as $n $
quadratic in momenta $\mu$ Hamiltonians on the phase space $\mathcal{M}%
=T^{\ast}\mathcal{Q}$ cotangent to a Riemannian manifold $\mathcal{Q}$
parametrized by $\left(  \lambda_{1},\ldots,\lambda_{n}\right)  $ and equipped
with the contravariant metric tensor $G^{(m)}$ (depending on $m\in\mathbf{Z}$)
given by:%
\begin{equation}
G^{(m)}=\operatorname*{diag}\left(  \frac{\lambda_{1}^{m}}{\Delta_{1}}%
,\ldots,\frac{\lambda_{n}^{m}}{\Delta_{n}}\right)  \text{ \ \ with }\Delta
_{i}=%
{\textstyle\prod\limits_{j\neq i}}
(\lambda_{i}-\lambda_{j}). \label{metryka}%
\end{equation}
It can be shown that $G^{(m)}$ is of zero curvature for $m=0,\ldots,n$ and
that $G^{(n+1)}$ is of non-zero constant curvature, while all other choices of
$m$ lead to spaces of non-constant curvature. The Hamiltonians $H_{i}^{n,m,k}$
are known in literature as St\"{a}ckel Hamiltonians and the corresponding
commuting Hamiltonian flows (\ref{2.2}) are then called St\"{a}ckel systems,
or more precisely, St\"{a}ckel systems of Benenti type. They are obviously
separable in the sense of Hamilton-Jacobi theory since they by the very
definition satisfy St\"{a}ckel relations (\ref{BenSR}). The objects $K_{i}$ in
(\ref{HamB}) are Killing tensors for any metric $G^{(m)}$ and are given by
\[
K_{i}=-\operatorname*{diag}\left(  \frac{\partial q_{i}}{\partial\lambda_{1}%
},\cdots,\frac{\partial q_{i}}{\partial\lambda_{n}}\right)  \text{
\ \ \ }i=1,\ldots,n,
\]
where $q_{i}=q_{i}(\lambda)$ are Vi\`{e}te polynomials (signed symmetric
polynomials) in $\lambda$:%
\begin{equation}
q_{i}(\lambda)=(-1)^{i}%
{\displaystyle\sum\limits_{1\leq s_{1}<s_{2}<\ldots<s_{i}\leq n}}
\lambda_{s_{1}}\ldots\lambda_{s_{i}}\text{, \ \ }i=1,\ldots,n \label{defq}%
\end{equation}
that can also be considered as new coordinates on \ the Riemannian manifold
$\mathcal{Q}$ (we will then refer to them as Vi\`{e}te coordinates). Notice
that $K_{i}$ do not depend on neither $m$ nor $k$. Finally, the potentials
$V_{i}^{(k)}$ can be constructed recursively \cite{maciej} by%

\begin{equation}
V_{i}^{(k+1)}=V_{i+1}^{(k)}-q_{i}V_{1}^{(k)},\text{ }k\in\mathbf{Z}\text{,
\ \ \ with }V_{i}^{(0)}=\delta_{in}, \label{wprost}%
\end{equation}
where we put $V_{i}^{(k)}=0$ for $i<0$ or $i>n$. The first potentials are
trivial: $V_{i}^{(k)}=\delta_{i,n-k}$ for $k=0,1,\ldots,n-1$. The first
nontrivial potentials are $V_{i}^{(n)}=-q_{i},$ For \thinspace$k>n$ the
potentials $V_{i}^{(k)}$ become complicated polynomial functions of $q$. The
recursion (\ref{wprost}) can also be reversed
\begin{equation}
V_{r}^{(k)}=V_{r-1}^{(k+1)}-\frac{q_{r-1}}{q_{n}}V_{n}^{(k+1)}\text{,}%
\ \ k\in\mathbf{Z}\text{, }r=1,\ldots,n\text{,} \label{rekdown}%
\end{equation}
leading to potentials $V_{i}^{(k)}$ with $k<0$. These potentials start with
$V_{i}^{(-1)}=-\frac{q_{i-1}}{q_{n}}$ and are rather complicated rational
functions of $q$. They will be referred to as negative potentials. It can also
be shown \cite{bensol} that
\begin{equation}
g_{ij}^{(m)}=V_{1}^{(2n-m-i-j)} \label{gij}%
\end{equation}
where $g^{(m)}=\left(  G^{(m)}\right)  ^{-1}$ is the corresponding covariant
metric tensor.

\begin{remark}
The general $n-$time (simultaneous) solution for Hamilton equations
(\ref{2.2}) associated with all the Hamiltonians (\ref{HamB})\ is given
implicitly by
\end{remark}

\begin{equation}
t_{i}+c_{i}=\pm\frac{1}{2}\sum_{r=1}^{n}%
{\displaystyle\int}
\frac{\lambda_{r}^{n-i}}{\sqrt{\lambda_{r}^{m}\left(
{\textstyle\sum\nolimits_{j=1}^{n}}
a_{j}\lambda_{r}^{n-j}-\frac{\varepsilon}{4}\lambda_{r}^{k}\right)  }}%
d\lambda_{r}\text{, \ \ \ }i=1,\ldots,n\text{.} \label{gensol}%
\end{equation}
To see this it is enough to integrate the related Hamilton-Jacobi problem.
Now, with $n$ Hamiltonians $H_{i}^{n,m,k}$ in (\ref{HamB}) we can associate,
by corresponding Legendre transforms, $n$ Lagrangians $L_{i}^{n,m,k}%
:T\mathcal{Q\rightarrow}\mathbf{R}$ given by%
\begin{equation}
L_{i}^{n,m,k}(\lambda,\lambda_{t_{i}})=\frac{1}{4}\lambda_{t_{i}}^{T}%
g^{(m)}K_{i}^{-1}\lambda_{t_{i}}-\frac{\varepsilon}{4}V_{i}^{(k)},\text{
\ }i=1,\ldots,n. \label{Li}%
\end{equation}
Every Lagrangian $L_{i}^{n,m,k}$ give rise to $n$ systems of Euler-Lagrange
equations
\begin{equation}
E_{j}^{s}(L_{i}^{n,m,k})=0\text{, \ \ }j=1,\ldots,n \label{EL}%
\end{equation}
(each for every $s$ between $1$ and $n$) where
\[
E_{j}^{s}=\frac{\partial}{\partial\lambda_{j}}-\frac{d}{dt_{s}}\frac{\partial
}{\partial\left(  \partial\lambda_{j}/\partial t_{s}\right)  }\text{,
\ \ \ }j=1,\ldots,n
\]
are components of the Euler-Lagrange operator with respect to the independent
variable $t_{a}$.

\begin{remark}
By construction, the solutions (\ref{gensol}) are also general solutions for
all the Euler-Lagrange equations (\ref{EL}). It means that for a particular
$s$ the general solution of Euler-Lagrange equations $E_{j}^{s}(L_{i}%
^{n,m,k})=0$ is given by (\ref{gensol}) where $t_{p}$ for $p\neq s$ plays a
role of a constant parameter.
\end{remark}

Denote now the variable $t_{1}$ as $x$ (our method works similarly with any
$t_{i}$ chosen as $x$). With every Killing tensor $K_{i}$ for $i=2,\ldots,n$
we can associate a dispersionless evolutionary PDE of the form%
\begin{equation}
\lambda_{t_{i}}=K_{i}\lambda_{x}\equiv Z_{i}\left[  \lambda\right]  \text{
\ \ \ }i=2,\ldots,n \label{Kilsys}%
\end{equation}
(where $\lambda=\left(  \lambda_{1},\ldots\lambda_{n}\right)  ^{T}$). We will
call PDE's in (\ref{Kilsys}) simply\emph{\ Killing systems. }Here and in what
follows we use the notation\emph{\ }$f\left[  \lambda\right]  $ to denote
integral-differential function of $\lambda$ i.e. a function of $\lambda$, its
$x$-derivatives and antiderivatives (integrals). In the case above
$Z_{i}\left[  \lambda\right]  =Z_{i}(\lambda,\lambda_{x})$. The chosen
variable $t_{1}=x$ in (\ref{Kilsys}) plays thus the role of a space variable
while the remaining variables $t_{i}$ should then be considered as evolution
parameters (times). Equations (\ref{Kilsys}) constitute a set of $n-1$
integrable dispersionless equations that due to the form of $K_{i}$ belong to
the class of weakly nonlinear semi-Hamiltonian systems, i.e. hydrodynamic-type
systems that are semi-Hamiltonian in the sense of Tsarev \cite{tsar1}%
,\cite{tsar2} and weakly nonlinear \cite{ferap1}. Actually, the systems
(\ref{Kilsys}) are finite-component restrictions of the universal hydrodynamic
hierarchy considered in \cite{al1}. The variables $\lambda_{i}$ are Riemann
invariants of all the system (\ref{Kilsys}) as $K_{i}$ are diagonal in
$\lambda$. The systems (\ref{Kilsys}) can also be considered as $n-1$
dynamical systems on some infinite-dimensional function space $\mathcal{V}$ of
vectors $(\lambda_{1}(x),\ldots,\lambda_{n}(x))$. with $Z_{i}$ being $n-1$
vector fields on $\mathcal{M}$. It can be shown \cite{ferap1} that the vector
fields $Z_{i}$ commute on $\mathcal{V}$:%
\[
\left[  Z_{i},Z_{j}\right]  =0\ \ \ \ \ \ i,j=2,\ldots,n.
\]
Note also that since $K_{1}=I$ we can complete the system of equations
(\ref{Kilsys}) by the equation $\lambda_{\tau}=K_{1}\lambda_{x}=\lambda
_{x}\equiv Z_{1}$with the translation-invariant general solution $\lambda
_{i}=\lambda_{i}(x+\tau)$. The vector field $Z_{1}$ also commutes with all the
vector fields $Z_{2},\ldots,Z_{n}$ \cite{ferap1}.

\begin{proposition}
Every mutual solution $\lambda(t_{1},\ldots,t_{n})$ (\ref{gensol}) of all
Hamiltonian systems (\ref{2.2}) with Hamiltonians of Benenti type (\ref{HamB})
is (after replacing $t_{1}$ with $x$) also a particular solution of all $n-1$
corresponding Killing systems in (\ref{Kilsys}).
\end{proposition}

\begin{proof}
Let us assume that a vector function $\lambda(t_{1},\ldots,t_{n})$ solves
(\ref{gensol}). Then, by construction, it also solves the spatial part of
(\ref{2.2}) with appropriate functions $\mu(t_{1},\ldots,t_{n})$ given by
$\mu_{i}=\partial W(\lambda,a)/\partial\lambda_{i}$ ($W=W(\lambda,a)$ is a
common integral of all the Hamilton-Jacobi equations for Hamiltonians
$H_{i}^{n,m,k}$). It means that $\lambda(t_{1},\ldots,t_{n})$ solves
\begin{equation}
\lambda_{t_{i}}=\frac{\partial}{\partial\mu}H_{i}^{n,m,k}=2K_{i}G^{(m)}%
\mu,\ \ \ \ \ \ i=1,...,n. \label{eliminuj}%
\end{equation}
Since $K_{1}=I$ we get from the first equation in (\ref{eliminuj}) $\mu
(t_{1},\ldots,t_{n})=\frac{1}{2}g^{(m)}(\lambda(t_{1},\ldots,t_{n}%
))\lambda_{t_{1}}(t_{1},\ldots,t_{n})$. Substituting it to the remaining
equations in (\ref{eliminuj}) yields then
\[
\lambda_{t_{i}}(t_{1},\ldots,t_{n})=K_{i}(\lambda(t_{1},\ldots,t_{n}%
))\lambda_{t_{1}}(t_{1},\ldots,t_{n})\text{, \ \ \ \ }i=2,\ldots,n
\]
which concludes the proof as $t_{1}=x$. Thus, all the solutions (\ref{gensol})
also solve all $n-1$ Killing systems in (\ref{Kilsys}).
\end{proof}

Moreover, we have

\begin{theorem}
The general ($n$-time) solution of all the Killing systems in (\ref{Kilsys})
is given by%
\begin{equation}
t_{i}+c_{i}=\sum_{r=1}^{n}%
{\displaystyle\int}
\frac{\lambda_{r}^{n-i}}{\varphi_{r}(\lambda_{r})}d\lambda_{r}%
,\ \ \ \ i=1,...,n \label{genKil}%
\end{equation}
(where $\varphi_{r}$ are arbitrary functions of one variable)
\end{theorem}

The proof of this statement can be found in \cite{ferap1}. Obviously,
(\ref{genKil}) contains all the solutions (\ref{gensol}).

Suppose now that a particular solution (\ref{genKil}) of our Killing systems
(\ref{Kilsys}) is of the more specific form (\ref{gensol}). Since this class
of solutions - by construction - satisfies all the Euler-Lagrange equations
(\ref{EL}), we can treat equations (\ref{EL}) as additional bonds that these
solutions satisfy. We can therefore use these bonds to express some variables
$\lambda_{i}$ by other $\lambda$'s. Thus, within the class (\ref{gensol}) of
solutions (\ref{genKil}) of Killing systems (\ref{Kilsys}) we can perform a
\emph{variable elimination} (reparametrization) that turns (\ref{Kilsys}) into
entirely new sets of evolutionary PDE's. \ As we have demonstrated in
\cite{bensol} and in \cite{makrzkdv}, in carefully chosen cases and in a
particular coordinate system (Vi\`{e}te coordinates (\ref{defq})) this
reparametrization turns systems (\ref{Kilsys}) into systems with dispersion
(soliton hierarchies) with the solution (\ref{gensol}) being also a solution
of these new systems with dispersion. In this paper we will produce by this
method (the local and the nonlocal part of) the coupled (multicomponent) Harry
Dym hierarchy.

\section{Nonlocal coupled Harry Dym hierarchy}

Assume now that $\varepsilon=1$ in (\ref{BenSR}) and therefore also in
(\ref{gensol}), (\ref{Li}) etc.). In order to perform the elimination
procedure just mentioned, let us pass to Vi\`{e}te coordinates as given in
(\ref{defq}). The Killing systems (\ref{Kilsys}) are tensorial so in Vi\`{e}te
coordinates they have the form
\[
q_{t_{i}}=K_{i}(q)q_{x},\ \ \ \ i=2,\ldots,n
\]
or, explicitly
\begin{equation}
\frac{d}{dt_{i}}q_{j}=(q_{j+i-1})_{x}+\sum_{k=1}^{j-1}\left(  q_{k}\left(
q_{j+i-k-1}\right)  _{x}-q_{j+i-k-1}\left(  q_{k}\right)  _{x}\right)
\equiv\left(  Z_{i}^{n}\left[  q\right]  \right)  ^{j},\text{\ \ \ }%
j=1,\ldots,n \label{Kilq}%
\end{equation}
(where we put $q_{\alpha}=0$ for $\alpha>n$), where $i=2,\ldots,n$ and where
$\left(  Z_{i}^{n}\left[  q\right]  \right)  ^{j}$ denotes the $j$-th
component of the vector field $Z_{i}\left[  q\right]  $. The superscript $n$
at $Z_{i}$ indicates the number of components in the vector field $Z_{i}$ and
we will sometimes use it since we will need to switch between various $n$.
From (\ref{Kilq}) one can see that $\left(  Z_{i}^{n}\left[  q\right]
\right)  ^{j}=\left(  Z_{j}^{n}\left[  q\right]  \right)  ^{i}$ for all
$i,j=1,\ldots,n$. Obviously, $G^{(m)}$, $g^{(m)}$ and $K_{i}$ are tensors and
can thus also easily be transformed to Vi\`{e}te coordinates.

Consider now Euler-Lagrange equations (\ref{EL}) with $s=1$ (so that
$t_{s}=t_{1}=x$) associated with Lagrangians $L_{1}^{n,m,k}$ denoted further
on for simplicity as $L^{n,m,k}$. Denote also $E_{i}^{1}$ as $E_{i}$,
$i=1,\ldots,n$ and consider the equations%
\begin{equation}
E_{i}\left(  L^{n,m,k}\right)  =0\text{, }i=1,\ldots,n\text{, \ \ \ }%
n,\in\mathbf{N},\text{ }k\in\mathbf{Z,} \label{ELspec}%
\end{equation}
written in $q$-variables, so that now
\[
E_{i}=\frac{\partial}{\partial q_{i}}-\frac{d}{dx}\frac{\partial}{\partial
q_{i,x}},\ \ \ i=1,\ldots,n,
\]
while (since $K_{1}=I$)%
\begin{equation}
L^{n,m,k}=L^{n,m,k}(q,q_{x})=\frac{1}{4}q_{x}^{T}g^{(m)}q_{x}-\frac{1}{4}%
V_{1}^{(k)}. \label{L1}%
\end{equation}
As it has been shown in \cite{bensol} the following symmetry relations are
satisfied for $\alpha=1,\ldots,n-1$%
\begin{equation}
E_{i}\left(  L^{n,m,k}\right)  =E_{i-\alpha}\left(  L^{n,m+\alpha,k-\alpha
}\right)  ,\ \ \ \ \ \ \ i=\alpha+1,...,n, \label{wtyl}%
\end{equation}
that can also be written as%
\begin{equation}
E_{i}\left(  L^{n,m,k}\right)  =E_{i+\alpha}\left(  L^{n,m-\alpha,k+\alpha
}\right)  ,\ \ \ \ \ \ \ i=1,...,n-\alpha. \label{wprzod}%
\end{equation}
Due to (\ref{wtyl}) and (\ref{wprzod}) the equations (\ref{ELspec}) can be
embedded in the following double-infinite multi-Lagrangian "ladder" of
Euler-Lagrange equations of the form%
\begin{equation}
E_{1}\left(  L^{n,m+j-1,k-j+1}\right)  =E_{2}\left(  L^{n,m+j-2,k-j+2}\right)
=\cdots=E_{n}(L^{n,m+j-n,k-j+n})=0\text{, \ }j=\ldots,-1,0,1,\ldots
\label{ladder}%
\end{equation}
with fixed $m,k\in\mathbf{Z}$ (the equations (\ref{ELspec}) fit in
(\ref{ladder}) at $j=1,2,\ldots,n$). For a given dimension $n$ the ladder
(\ref{ladder}) is determined by the sum $m+k$ in the sense that various
choices of $m$ and $k$ with the same $m+k$ yield the same ladder.

We are now ready to present our elimination procedure leading to
multicomponent integral (nonlocal) Harry Dym hierarchy. Assume that we want to
produce first $s-1$ \ flows of the $N$-component ($N\in\mathbf{N}$) hierarchy.
Let us take $n=s+N-1$, $m=-N$ and $k=0$ in (\ref{Li}), that is, let us
consider the purely kinetic Lagrangian $L^{n,-N,0}$ with $n=s+N-1$ and the
corresponding Euler-Lagrange equations (\ref{ELspec}). Due to this special
choice of all parameters \emph{the last} $n-N$ equations in (\ref{ELspec})
attain the form%
\begin{equation}%
\begin{array}
[c]{l}%
E_{N+1}\left(  L^{n,-N,0}\right)  \equiv-\frac{1}{2}q_{n,xx}+\varphi
_{n-N}[q_{1},...,q_{n-1}]=0,\\
E_{N+2}\left(  L^{n,-N,0}\right)  \equiv-\frac{1}{2}q_{n-1,xx}+\varphi
_{n-N-1}[q_{1},...,q_{n-2}]=0,\\
\vdots\\
E_{n-1}\left(  L^{n,-N,0}\right)  \equiv-\frac{1}{2}q_{N+2,xx}+\varphi
_{2}[q_{1},...,q_{N+1}]=0\\
E_{n}\left(  L^{n,-N,0}\right)  \equiv-\frac{1}{2}q_{N+1,xx}+\varphi_{1}%
[q_{1},...,q_{N}]=0.
\end{array}
\label{EL1}%
\end{equation}
and are a part of the ladder (\ref{ladder}) with $m+k=-N$. Now, by direct
calculation of $E_{i}\left(  L^{n,-N,0}\right)  $ with the use of some
identities satisfied by the potentials $V_{1}^{(i)}$ it can be proved that%
\begin{align*}
E_{N}\left(  L^{n,-N,0}\right)   &  =E_{N+1}\left(  L^{n+1,-N,0}\right)
+\frac{1}{2}q_{n+1,xx},\\
E_{i}\left(  L^{n,-N,0}\right)   &  =E_{i+1}\left(  L^{n+1,-N,0}\right)
\text{, \ \ }i=N+1,\ldots,n.
\end{align*}
These identities lead to

\begin{proposition}
\label{niezalezy}The functions $\varphi_{i}$ in (\ref{EL1}) do not depend on
$n$ in the sense that increasing $n$ to $n+1$ (and keeping $N$ constant) turn
(\ref{EL1}) into $n-N+1$ equations
\begin{equation}%
\begin{array}
[c]{llllll}%
E_{N+1}\left(  L^{n+1,-N,0}\right)  & = & -\frac{1}{2}q_{n+1,xx}+E_{N}\left(
L^{n,-N,0}\right)  & \equiv & -\frac{1}{2}q_{n+1,xx}+\varphi_{n-N+1}%
[q_{1},...,q_{n}] & =0,\\
E_{N+2}\left(  L^{n+1,-N,0}\right)  & = & E_{N+1}\left(  L^{n,-N,0}\right)  &
\equiv & -\frac{1}{2}q_{n,xx}+\varphi_{n-N}[q_{1},...,q_{n-1}] & =0,\\
\vdots &  & \vdots &  & \vdots & \\
E_{n}\left(  L^{n+1,-N,0}\right)  & = & E_{n-1}\left(  L^{n,-N,0}\right)  &
\equiv & -\frac{1}{2}q_{N+2,xx}+\varphi_{2}[q_{1},...,q_{N+1}] & =0,\\
E_{n+1}\left(  L^{n+1,-N,0}\right)  & = & E_{n}\left(  L^{n,-N,0}\right)  &
\equiv & -\frac{1}{2}q_{N+1,xx}+\varphi_{1}[q_{1},...,q_{N}] & =0.
\end{array}
\label{EL2}%
\end{equation}

\end{proposition}

It means that increasing $n$ to $n+1$ (and keeping $N$ constant) in
(\ref{EL1}) des not alter these equations except that a new equation of the
form
\[
E_{N+1}\left(  L^{n+1,-N,0}\right)  \equiv-\frac{1}{2}q_{n+1,xx}%
+\varphi_{n-N+1}[q_{1},...,q_{n}]=0
\]
is added at the top of (\ref{EL1}). As we will see soon, this will result in
the fact that our construction indeed yields an infinite hierarchy of
commuting flows.

Due to their structure, equations (\ref{EL1}) can be formally solved with
respect to the variables $q_{N+1},\ldots,q_{n},$ which yields $q_{N+1}%
,\ldots,q_{n}$ as some nonlocal (integral-differential) functions of
$q_{1},\ldots,q_{N}$:%

\begin{equation}%
\begin{array}
[c]{l}%
q_{N+1}=f_{1}\left[  q_{1},\ldots,q_{N}\right] \\
\vdots\\
q_{n}=f_{n-N+1}\left[  q_{1},\ldots,q_{N}\right]  ,
\end{array}
\label{eli2}%
\end{equation}
where, due to Proposition \ref{niezalezy}, the functions $f_{i}$ do not depend
on $n,$ so increasing $n$ by $1$ (and keeping $N$ constant) will only result
in one new equation at the bottom place in (\ref{eli2}). Let us now replace
the variables $q_{N+1},\ldots,q_{n}$ in the first $N$ components of the first
$s-1$ Killing systems (\ref{Kilq}) by the corresponding functions $f_{i}$
(right-hand sides of (\ref{eli2})). This yields equations of the form%
\begin{equation}
\overline{q}_{t_{r}}=\overline{Z}_{r}^{N}\left[  \overline{q}\right]
\text{\ \ \ \ }r=2,\ldots s \label{firsts}%
\end{equation}
where $\overline{q}$ denotes the first $N$ entries in $q$ i.e. $\overline
{q}=\left(  q_{1},\ldots,q_{N}\right)  ^{T}$. They are in general highly
nonlinear autonomous systems of $N$ evolution equations for $q_{1}%
,\ldots,q_{N}.$

\begin{theorem}
\label{incrs}The vector fields $\overline{Z}_{r}^{N}\left[  \overline
{q}\right]  $ in (\ref{firsts}) do not depend on $s$ in the sense that if we
increase $s$ by one in our procedure then (\ref{firsts}) are unaltered and a
new equation $\overline{q}_{t_{s+1}}=\overline{Z}_{s+1}^{N}\left[
\overline{q}\right]  $ appears.
\end{theorem}

\begin{proof}
This theorem is a consequence of Proposition \ref{niezalezy}. If we increase
$s$ to $s+1$ and keep $N$ constant we have to take $n+1$ instead of $n$ in our
procedure as $n=s+N-1$. Due to (\ref{Kilq}) we have \ $\left(  Z_{r}%
^{n+1}\right)  ^{j}=\left(  Z_{r}^{n}\right)  ^{j}$ for $r=2,\ldots,s$ and for
$j=1,\ldots,N\,\ $i.e. the first $N$ components of the first $s-1$ of Killing
systems (\ref{Kilq}) do not change when we increase $n$ to $n+1$. Moreover, as
we explained above, the $n-N$ functions $f_{i}$ in (\ref{eli2}) do not change
either. So, the elimination procedure for the first $s-1$ vector fields
$Z_{i}$ is not altered leading to exactly the same vector fields $\overline
{Z}_{r}^{N}\left[  \overline{q}\right]  $ with $r=2,\ldots,s$ while the vector
field $Z_{s+1}^{n+1}$ yields the vector field $\overline{Z}_{s+1}^{N}\left[
\overline{q}\right]  $ i.e. a new equation at the end of the sequence
(\ref{firsts}).
\end{proof}

Repeating this argument we can increase $s$ indefinitely. Thus, our procedure
leads to an infinite hierarchy of evolutionary vector fields (flows)
\begin{equation}
\overline{q}_{t_{r}}=\overline{Z}_{r}^{N}\left[  \overline{q}\right]
\text{\ \ \ \ }r=2,3,\ldots\label{cel1}%
\end{equation}
in the sense that if we wish to produce any first $s-1$ flows (\ref{firsts})
of the hierarchy we can perform our procedure with $n=s+N-1$. This way we can
obtain arbitrary long sequences of the same infinite set of vector fields with
dispersion that pairwise commute (soliton hierarchy):

\begin{theorem}
\label{komm}The vector fields $\overline{Z}_{r}^{N}\left[  \overline
{q}\right]  $ commute i.e.%
\[
\left[  \overline{Z}_{i}^{N},\overline{Z}_{j}^{N}\right]  =0\text{ for any
}i,j=2,3,\ldots.
\]

\end{theorem}

This theorem is due to the fact that the original vector fields $Z_{i}^{n}$
commute and that the Euler-Lagrange equations $E_{i}(L^{(n,m,k)})=0$ are
invariant with respect to all the fields $Z_{i}^{n}$ \cite{bensol}. Moreover,
the vector fields $\overline{Z}_{i}^{N}$ still commute with $\overline{Z}%
_{1}^{N}=\left(  q_{1,x},\ldots,q_{N,x}\right)  ^{T}$. As we demonstrate
below, the hierarchy (\ref{cel1}) is the nonlocal part of the multicomponent
Harry Dym soliton hierarchy as discussed in \cite{brunelli}.

\begin{example}
Consider first $N=1$ (one-component hierarchy as discussed in \cite{brunelli}%
). Suppose that we want to obtain the first $s-1=2$ flows of the hierarchy. We
have then to take $n=s+N-1=3$ and consider the elimination equations
(\ref{EL1}) for these parameters. The pure kinetic Lagrangian $L^{3,-1,0}$ has
the form%
\[
L^{3,-1,0}=\frac{1}{2}q_{1,x}^{2}\left(  q_{1}q_{2}-\frac{1}{2}q_{3}-\frac
{1}{2}q_{1}^{3}\right)  +\frac{1}{2}q_{1,x}q_{2,x}\left(  q_{1}^{2}%
-q_{2}\right)  -\frac{1}{4}q_{2,x}^{2}q_{1}-\frac{1}{2}q_{1,x}q_{3,x}%
q_{1}+\frac{1}{2}q_{2,x}q_{3,x}%
\]
so that (\ref{EL}) become%
\begin{align}
E_{2}\left(  L^{3,-1,0}\right)   &  \equiv-\frac{1}{2}q_{3,xx}+\frac{1}%
{2}q_{1,xx}q_{2}+\frac{1}{2}q_{2,xx}q_{1}-\frac{1}{2}q_{1,xx}q_{1}^{2}%
-\frac{1}{2}q_{1}q_{1,x}^{2}+\frac{1}{2}q_{1,x}q_{2,x}=0,\label{el}\\
E_{3}\left(  L^{3,-1,0}\right)   &  \equiv-\frac{1}{2}q_{2,xx}+\frac{1}%
{2}q_{1,xx}q_{1}+\frac{1}{4}q_{1,x}^{2}=0.\nonumber
\end{align}
Due to their specific structure, we can solve (\ref{el}) with respect to
$q_{2}$ and $q_{3}$. We will thus use (\ref{el}) to eliminate variables in the
corresponding $n=3$-component Killing systems (\ref{Kilq}) that have in this
case the form:
\begin{align}
\frac{d}{dt_{2}}\left(
\begin{array}
[c]{c}%
q_{1}\\
q_{2}\\
q_{3}%
\end{array}
\right)   &  =\left(
\begin{array}
[c]{c}%
q_{2,x}\\
q_{3,x}+q_{1}q_{2,x}-q_{2}q_{1,x}\\
q_{1}q_{3,x}-q_{3}q_{1,x}%
\end{array}
\right)  =Z_{2}^{3},\nonumber\\
\frac{d}{dt_{3}}\left(
\begin{array}
[c]{c}%
q_{1}\\
q_{2}\\
q_{3}%
\end{array}
\right)   &  =\left(
\begin{array}
[c]{c}%
q_{3,x}\\
q_{1}q_{3,x}-q_{3}q_{1,x}\\
q_{2}q_{3,x}-q_{3}q_{2,x}%
\end{array}
\right)  =Z_{3}^{3}. \label{ela}%
\end{align}
By the second equation in (\ref{el}) we obtain%
\[
q_{2,xx}=\frac{1}{2}q_{1,x}^{2}+q_{1,xx}q_{1}.
\]
Integrating it once we obtain%
\[
q_{2,x}=\frac{1}{2}q_{1}q_{1,x}+\frac{1}{2}\partial^{-1}q_{1}q_{1,xx}%
\]
where
\[
\partial^{-1}=\int\ldots dx+\varphi(t_{2},t_{3})
\]
is the integration operator with the integration parameter $\varphi$ that has
to be chosen from case to case and has therefore to be treated as a part of
the solution of every integration problem. It is always possible to find such
a function. Integrating $q_{2,x}$ we obtain%
\[
q_{2}=\frac{1}{4}q_{1}^{2}+\frac{1}{2}\partial^{-2}q_{1}q_{1,xx}.
\]
Further, the first equation in (\ref{el}) yields%
\begin{equation}
q_{3,xx}=q_{1,xx}q_{2}+q_{2,xx}q_{1}-q_{1,xx}q_{1}^{2}-q_{1}q_{1,x}%
^{2}+q_{1,x}q_{2,x} \label{q3xx}%
\end{equation}
Inserting to it $q_{2}$ and $q_{2,x},$ as calculated above, and integrating
once we obtain%
\begin{equation}
q_{3,x}=-\frac{1}{2}\partial^{-1}q_{1}^{2}q_{1,xx}+\frac{1}{2}q_{1}%
\partial^{-1}q_{1}q_{1,xx}+\frac{1}{4}q_{1}^{2}q_{1,x}+\frac{1}{2}%
q_{1,x}\partial^{-2}q_{1}q_{1,xx} \label{q3x}%
\end{equation}
By inserting the obtained formulas for $q_{2,x}$ and $q_{3,x}$ into the first
$N=1$ components of $Z_{2}$ and $Z_{3}$ we obtain the first two flows of our
nonlocal soliton hierarchy:%
\begin{align}
q_{1,t_{2}}  &  =\frac{1}{2}q_{1}q_{1,x}+\frac{1}{2}\partial^{-1}q_{1}%
q_{1,xx}=\overline{Z}_{2},\label{poletka}\\
q_{1,t_{3}}  &  =-\frac{1}{2}\partial^{-1}q_{1}^{2}q_{1,xx}+\frac{1}{2}%
q_{1}\partial^{-1}q_{1}q_{1,xx}+\frac{1}{4}q_{1}^{2}q_{1,x}+\frac{1}{2}%
q_{1,x}\partial^{-2}q_{1}q_{1,xx}=\overline{Z}_{3}.\nonumber
\end{align}
Observe that in this particular case we did not have to calculate $q_{3}$
since it does not enter into the first component of neither $Z_{2}$ nor
$Z_{3}$. We needed however $q_{2}$ in order to calculate $q_{3,x}$. The flows
(\ref{poletka}) commute due to Theorem \ref{komm}.
\end{example}

\begin{example}
Let us now take $N=2$ and $s-1=1$ so that $n=3$ again. We will thus eliminate
$n-N=1$ variables (namely $q_{3}$) from the first $N=2$ components of the
field $Z_{2}^{3}$ above. The elimination equations (\ref{EL}) reduce now to
$E_{3}\left(  L^{3,-2,0}\right)  =0$. But, according to (\ref{wtyl}),
$E_{3}\left(  L^{3,-2,0}\right)  =E_{2}\left(  L^{3,-1,-1}\right)
=E_{2}\left(  L^{3,-1,0}\right)  $, the last equality due to the fact that
$L^{3,-1,-1}=L^{3,-1,0}-\frac{1}{4}V_{1}^{(-1)}=L^{3,-1,0}+\frac{1}{4q_{3}}$.
Thus, the elimination equation $E_{3}\left(  L^{3,-2,0}\right)  =0$ coincides
with the first equation in (\ref{el}) and yields exactly (\ref{q3xx}).
Plugging its integrated form (\ref{q3x}) into the first two components of
$Z_{2}^{3}$ yields the first flow of the $2$-component nonlocal cHD hierarchy:%
\begin{equation}
\frac{d}{dt_{2}}\left(
\begin{array}
[c]{c}%
q_{1}\\
q_{2}%
\end{array}
\right)  =\left(
\begin{array}
[c]{c}%
q_{2,x}\\
q_{1}q_{2,x}-q_{2}q_{1,x}-\frac{1}{2}\partial^{-1}q_{1}^{2}q_{1,xx}+\frac
{1}{2}q_{1}\partial^{-1}q_{1}q_{1,xx}+\frac{1}{4}q_{1}^{2}q_{1,x}+\frac{1}%
{2}q_{1,x}\partial^{-2}q_{1}q_{1,xx}%
\end{array}
\right)  =\overline{Z}_{2} \label{dwupolowy}%
\end{equation}

\end{example}

The map%
\begin{equation}
u_{i}=E_{N-i+1}\left(  L^{N,0,0}\right)  \text{, \ \thinspace}i=1,\cdots,N
\label{mapka}%
\end{equation}
transforms the hierarchy (\ref{cel1}) into the nonlocal part of the coupled
Harry Dym hierarchy (see \cite{AF2} for its local part) that is the
generalization of the one-field nonlocal HD hierarchy presented in
\cite{brunelli}. For example, for $N=2$ this map reads%
\begin{align*}
u_{1}  &  =-\frac{1}{2}q_{1,xx}\\
u_{2}  &  =-\frac{1}{2}q_{2,xx}+\frac{1}{4}q_{1,x}^{2}+\frac{1}{2}%
q_{1}q_{1,xx}%
\end{align*}
and applied to the field $\overline{Z}_{2}$ above yields%
\[
\frac{d}{dt_{2}}\left(
\begin{array}
[c]{c}%
u_{1}\\
u_{2}%
\end{array}
\right)  =-2\left(
\begin{array}
[c]{c}%
u_{1,x}\partial^{-2}u_{1}+2u_{1}\partial^{-1}u_{1}-\frac{1}{2}u_{2,x}\\
u_{2,x}\partial^{-2}u_{1}+2u_{2}\partial^{-1}u_{1}%
\end{array}
\right)  .
\]

\section{Solutions of the multicomponent nonlocal HD hierarchy}

We will now construct a variety of solutions of the hierarchy (\ref{cel1}).

\begin{theorem}
\label{main}For any $\beta\in\{0,\ldots,n-1\}$ the functions $\lambda
_{i}=\lambda_{i}(t_{1},\ldots,t_{n})$ given implicitly by%
\begin{equation}
t_{i}+c_{i}=\pm\frac{1}{2}\sum_{r=1}^{n}%
{\displaystyle\int}
\frac{\lambda_{r}^{n-i}}{\sqrt{\lambda_{r}^{-N+\beta}\left(
{\textstyle\sum\nolimits_{j=1}^{n}}
a_{j}\lambda_{r}^{n-j}-\frac{1}{4}\lambda_{r}^{-\beta}\right)  }}d\lambda
_{r}\text{, \ \ \ }i=1,\ldots,n\text{.} \label{solsintcHD}%
\end{equation}
are such that the corresponding functions \thinspace$q_{i}=q_{i}(x=t_{1}%
,t_{2},\ldots,t_{n}),i=1,\ldots,N$, given by (\ref{defq}) are solutions of the
first $n-\beta$ ($n-1$ for $\beta=0,1$) equations of the $N$-component
integral cHD hierarchy (\ref{cel1}). The variables $t_{2},\ldots,t_{n-\beta
+1}$ ($t_{2},\ldots,t_{n}$ for $\beta=0,1$) play then the role of evolution
parameters (dynamical times) while the remaining $t_{i}$'s are free parameters.
\end{theorem}

For the proof of this theorem, see Appendix. We will now consider some
particular, interesting classes of solutions (\ref{solsintcHD}). Assume that
$\beta=0$ in (\ref{solsintcHD}) and that $a_{j}=\frac{1}{4}\delta_{j,n}%
+\delta_{j,n-\gamma}$ for some $\gamma\in\left\{  0,\ldots,n-1\right\}  $.
Then (\ref{solsintcHD}) attain the form%

\[
t_{i}+c_{i}=\pm\frac{1}{2}\sum_{r=1}^{n}%
{\displaystyle\int}
\frac{\lambda_{r}^{n-i}}{\sqrt{\lambda_{r}^{-N+\gamma}}}\,d\lambda_{r}\text{,
\ \ \ }i=1,\ldots,n\text{,}%
\]
that integrated yields%
\begin{equation}
t_{i}+c_{i}=\pm\frac{1}{2\left(  n-i+N/2-\gamma/2+1\right)  }\sum_{r=1}%
^{n}\lambda_{r}^{n-i+N/2-\gamma/2+1},\text{ \ \ \ }i=1,\ldots,n\text{.}
\label{ogint}%
\end{equation}
The above system can be algebraically solved with respect to $\lambda_{i}$
only for two choices of $\gamma$, namely $\gamma=N$ and $\gamma=N+1$, but it
turns out that the case $\gamma=N$ leads to trivial solutions (polynomial
solutions not depending on $x$). Thus, we must assume $\gamma=N+1$. In this
case the above equations attain the form%
\begin{equation}
t_{i}+c_{i}=\pm\frac{1}{2\left(  n-i+1/2\right)  }\sum_{r=1}^{n}\lambda
_{r}^{n-i+1/2},\text{ \ \ \ }i=1,\ldots,n\text{.} \label{int}%
\end{equation}
Note that (\ref{int}) do not depend on $N$. It means that for any $N$ between
$1$ and $n-2$ (as $\gamma=N+1\leq n-1$)the functions $q_{1}(x,t_{2}%
,\ldots,t_{n}),\ldots,q_{N}(x,t_{2},\ldots,t_{n})$ obtained from (\ref{int})
through (\ref{defq}) solve the first $n-1$ equations in (\ref{cel1}). The
following two examples illustrate this.

\begin{example}
Assume that $n=3$. Then (\ref{int}) attain the form (with $x=t_{1},$ $c_{i}=0
$, we also choose only $+$ in (\ref{int}))%
\begin{align}
x  &  =\frac{1}{5}%
{\textstyle\sum\nolimits_{i=1}^{3}}
z_{i}^{5}=\frac{1}{5}\left(  \rho_{1}^{5}-5(\rho_{1}\rho_{2}-\rho_{3}%
)(\rho_{1}^{2}-\rho_{2})\right) \nonumber\\
t_{2}  &  =\frac{1}{3}%
{\textstyle\sum\nolimits_{i=1}^{3}}
z_{i}^{3}=\frac{1}{3}\left(  \rho_{1}^{3}-3\rho_{1}\rho_{2}+3\rho_{3}\right)
\label{kasza}\\
t_{3}  &  =%
{\textstyle\sum\nolimits_{i=1}^{3}}
z_{i}=\rho_{1}\nonumber
\end{align}
where $z_{i}=\lambda_{i}^{1/2}$, $i=1,2,3$ and where $\rho_{1}=%
{\textstyle\sum\nolimits_{i=1}^{3}}
z_{i}$, $\rho_{2}=z_{1}z_{2}+z_{1}z_{3}+z_{2}z_{3}$ and $\rho_{3}=z_{1}%
z_{2}z_{3}$ are elementary symmetric polynomials in $z_{i}$. The right hand
sides of (\ref{kasza}) follow from Newton formulas:%
\begin{equation}%
{\textstyle\sum\nolimits_{i=1}^{n}}
z_{i}^{m}=%
{\displaystyle\sum\limits_{\alpha_{1}+2\alpha_{2}+\ldots+n\alpha_{n}=m}}
(-1)^{a_{2}+\alpha_{4}+\alpha_{6}+\cdots}m\frac{\left(  \alpha_{1}+\alpha
_{2}+\cdots+\alpha_{n}-1\right)  !}{\alpha_{1}!\ldots\alpha_{n}!}\rho
_{1}^{\alpha_{1}}\rho_{2}^{\alpha_{2}}\ldots\rho_{n}^{\alpha_{n}}\text{
\ \ for }m<n, \label{Newton}%
\end{equation}
expressing sums of powers of variables as functions of their symmetric
polynomials (these formulas can easily be extended to the case $m\geq n$ by
taking $n^{\prime}=m$ and putting all $\rho_{n+1},\ldots,\rho_{m}$ equal to
zero). The system (\ref{kasza}) can be solved explicitly yielding the solution
(\ref{int}) in $\rho$-variables:%
\begin{align}
\rho_{1}  &  =t_{3}\nonumber\\
\rho_{2}  &  =\frac{-15x-2t_{3}^{5}+15t_{3}^{2}t_{2}}{5\left(  3t_{2}%
-t_{3}^{3}\right)  }\label{ccc}\\
\rho_{3}  &  =\frac{15t_{2}t_{3}^{3}+45t_{2}^{2}-t_{3}^{6}-45xt_{3}}{15\left(
3t_{2}-t_{3}^{3}\right)  }\nonumber
\end{align}
On the other hand, according with (\ref{defq}) and with (\ref{Newton})
\[
q_{1}=-(\lambda_{1}+\lambda_{2}+\lambda_{3})=-\left(  z_{1}^{2}+z_{2}%
^{2}+z_{3}^{2}\right)  =-(2\rho_{2}-\rho_{1}^{2}).
\]
Plugging (\ref{ccc}) into the above identity we obtain%
\begin{equation}
q_{1}(x,t_{2},t_{3})=q_{1}(\rho_{i}(x,t_{2},t_{3}))=\frac{t_{3}^{5}%
+15t_{3}^{3}t_{2}-30x}{5\left(  3t_{2}-t_{3}^{3}\right)  }. \label{q1rozw}%
\end{equation}
According to Theorem \ref{main}, the function $q_{1}(x,t_{2},t_{3})$ given by
(\ref{q1rozw}) yield a two-time solution to the first $n-1=2$ flows of the
nonlocal 1-field (i.e. with $N=1$) HD hierarchy (\ref{cel1}), i.e. to both
systems (\ref{poletka}) (after an appropriate choice of integration constants).
\end{example}

\begin{example}
Let us now take $n=4$. In this case the equations (\ref{int}) read (again wit
all $c_{i}=0$ and with $+$ only and due to (\ref{Newton}))%
\begin{align}
x  &  =\frac{1}{7}%
{\textstyle\sum\nolimits_{i=1}^{4}}
z_{i}^{7}=\frac{1}{7}\left(  \rho_{1}^{7}-7(\rho_{1}\rho_{2}-\rho_{3})\left(
(\rho_{1}^{2}-\rho_{2})^{2}+\rho_{1}\rho_{3}\right)  -7\rho_{4}(\rho_{1}%
^{3}-2\rho_{1}\rho_{2}+\rho_{3})\right) \nonumber\\
t_{2}  &  =\frac{1}{5}%
{\textstyle\sum\nolimits_{i=1}^{4}}
z_{i}^{5}=\frac{1}{5}\left(  \rho_{1}^{5}-5(\rho_{1}\rho_{2}-\rho_{3}%
)(\rho_{1}^{2}-\rho_{2})-5\rho_{1}\rho_{4}\right) \label{kasza3}\\
t_{3}  &  =\frac{1}{3}%
{\textstyle\sum\nolimits_{i=1}^{4}}
z_{i}^{3}=\frac{1}{3}\left(  \rho_{1}^{3}-3\rho_{1}\rho_{2}+3\rho_{3}\right)
\nonumber\\
t_{4}  &  =%
{\textstyle\sum\nolimits_{i=1}^{4}}
z_{i}=\rho_{1}\nonumber
\end{align}
where as before $z_{i}=\lambda_{i}^{1/2}$ and $\rho_{i}$ are again symmetric
polynomials of the variables $z_{1},\ldots,z_{4}$. This system can again be
algebraically solved for $\rho_{1},\ldots,\rho_{4}$ although the solutions are
too complicated to present them here. We have now, according with
(\ref{defq}),%
\begin{align*}
q_{1}  &  =-(\lambda_{1}+\lambda_{2}+\lambda_{3}+\lambda_{4})=-\left(
z_{1}^{2}+z_{2}^{2}+z_{3}^{2}+z_{4}^{2}\right)  =-(2\rho_{2}-\rho_{1}^{2})\\
q_{2}  &  =\lambda_{1}\lambda_{2}+\cdots+\lambda_{3}\lambda_{4}=z_{1}^{2}%
z_{2}^{2}+\cdots+z_{3}^{2}z_{4}^{2}=\rho_{2}^{2}-2\rho_{1}\rho_{3}+2\rho_{4}%
\end{align*}
Substituting the variables $\rho_{i}$ obtained by solving (\ref{kasza3}) into
these expressions we obtain expressions for $q_{1}(x,t_{2},t_{3},t_{4})$ and
$q_{1}(x,t_{2},t_{3},t_{4})$:
\begin{equation}
q_{1}(x,t_{2},t_{3},t_{4})=\frac{P_{1}(x,t_{2},t_{3},t_{4})}{Q(t_{2}%
,t_{3},t_{4})}\text{, \ \ \ }q_{2}(x,t_{2},t_{3},t_{4})=\frac{P_{2}%
(x,t_{2},t_{3},t_{4})}{Q^{2}(t_{2},t_{3},t_{4})} \label{rozw}%
\end{equation}
where $P_{i}$ and $Q$ are rather complicated, but perfectly manageable for any
computer algebra program, polynomials. More specifically%
\[
P_{1}(x,t_{2},t_{3},t_{4})=-\frac{1}{7}\left(  105t_{4}^{3}t_{2}-t_{4}%
^{8}-21t_{4}^{5}t_{3}+630t_{2}t_{3}-630xt_{4}+315t_{3}^{2}t_{4}^{2}\right)
\]
and%
\[
Q(t_{2},t_{3},t_{4})=45t_{2}t_{4}+t_{4}^{6}-15t_{3}t_{4}^{3}-45t_{3}^{2},
\]
while $P_{2}$ is a quadratic in \thinspace$x$ polynom that is too complicated
to present it here. Now, according to Theorem \ref{main} and the theory above,
the function $q_{1}(x,t_{2},t_{3},t_{4})$ in (\ref{rozw}) solves the first
$n-1=3$ 1-field flows of the hierarchy (\ref{cel1}), so in particular both the
flows (\ref{poletka}), while the vector function%
\[
\left(
\begin{array}
[c]{c}%
q_{1}(x,t_{2},t_{3},t_{4})\\
q_{1}(x,t_{2},t_{3},t_{4})
\end{array}
\right)
\]
solves the first $n-1=3$ flows of the $N=2$-field cHD hierarchy (\ref{cel1})
starting with (\ref{dwupolowy}).
\end{example}

Let us also remark that formulas (\ref{ogint}) often lead to implicit
solutions of (\ref{cel1}). We illustrate it in the following example. Choose
$N=1$, $n=2$ and $\gamma=0$ in (\ref{ogint}). This yields (again for $c_{i}%
=0$)%
\begin{equation}
x=\frac{1}{5}\left(  z_{1}^{5}+z_{2}^{5}\right)  ,\text{ \ \ \ }t_{2}=\frac
{1}{5}\left(  z_{1}^{3}+z_{2}^{3}\right)  \label{1}%
\end{equation}
(with $z_{i}=\lambda_{i}^{1/2}$) that can not be algebraically solved.
However, (\ref{1}) can be embedded in the algebraically solvable system
(\ref{kasza}) in the sense that (\ref{kasza}) reduces to (\ref{1}) if we put
$z_{3}=0$ or equivalently $\rho_{3}=0,$ since $\rho_{3}=z_{1}z_{2}z_{3}$. By
virtue of Theorem \ref{main} it means that the function
\[
q_{1}(x,t_{2},y(x,t_{2}))=\frac{y(x,t_{2})^{5}+15y(x,t_{2})^{3}t_{2}%
-30x}{5\left(  3t_{2}-y(x,t_{2})^{3}\right)  }%
\]
with the variable $y(x,t_{2})$ defined implicitly by the equation
\[
15t_{2}y^{3}+45t_{2}^{2}-y^{6}-45xy=0
\]
(i.e. by the last equation in (\ref{ccc}) with $y$ instead of $t_{3}$), also
satisfies the first flow of the nonlocal HD-hierarchy i.e. the first flow in
(\ref{poletka}).

\section{Differential (local) cHD hierarchy and its solutions}

We will now obtain the purely differential part of cHD hierarchy as well as a
class of its implicit solutions. We choose now $\varepsilon=-1$ in
(\ref{BenSR}) in order to obtain real solutions in the local case (note that
it does not influence the potentials $V_{r}^{(m)}$). Analogously to the case
of nonlocal hierarchy, we will perform some variable elimination on the
sequence of Killing systems (\ref{Kilq}). Suppose thus that we want to produce
the first $s$ flows of the $N$-component local (i.e. purely differential)
Harry-Dym hierarchy. Put $n=s+N$ and consider the first $n-N$ Euler-Lagrange
equations for the Lagrangian $L^{n,n-N,-n}$. Using the fact that $V_{1}%
^{(-j)}=V_{1}^{(-j)}(q_{n-j+1},\ldots,q_{n})$ it can be shown that they attain
the form
\begin{align}
E_{1}\left(  L^{n,n-N,-n}\right)   &  \equiv\frac{1}{4q_{n}^{2}}+\,\gamma
_{1}^{\left(  N\right)  }[q_{1},\ldots,q_{N}]=0,\label{ELrozn}\\
E_{i}\left(  L^{n,n-N,-n}\right)   &  \equiv-\frac{q_{n-i+1}}{2q_{n}^{3}%
}\,+\gamma_{i,1}^{\left(  N\right)  }\left[  q_{1},\ldots,q_{N-i+1}\right]
+\frac{1}{q_{n}^{i+1}}\gamma_{i,2}^{(N)}\left[  q_{n-i+2},\ldots,q_{n}\right]
=0,\ \ i=2,...,n-N.\nonumber
\end{align}
where as usual $q_{\alpha}=0$ for $\alpha<1$. Note that (\ref{ELrozn}) and
(\ref{EL1}) belong to the same ladder (\ref{ladder}) of Euler-Lagrange
equations since in both cases $m+k=-N$.

\begin{proposition}
\label{nz2}The functions $\gamma_{i,1}^{\left(  N\right)  },\gamma
_{i,2}^{\left(  N\right)  },\gamma_{1}^{\left(  N\right)  }$ do not depend on
$n $ in the sense that increasing $n$ to $n+1$ will not alter (\ref{ELrozn})
except that a new equation originates at the bottom of the list (\ref{ELrozn}).
\end{proposition}

The proof of this proposition resembles the proof of the analogous statement
for nonlocal case i.e. Proposition \ref{niezalezy}. Note now that the
structure of (\ref{ELrozn}) makes it possible to eliminate (express) the
variables $q_{N+1},\ldots,q_{n}$ as (purely differential now) functions of
$q_{1},\ldots,q_{N}$ (although now, opposite to the nonlocal case, we first
calculate $q_{n}$, then $q_{n-1}$ and so on up to $q_{N+1}$):%
\begin{equation}%
\begin{array}
[c]{l}%
q_{n}=f_{1}^{(N)}\left[  q_{1},\ldots,q_{N}\right]  ,\\
\vdots\\
q_{N+1}=f_{n}^{(N)}\left[  q_{1},\ldots,q_{N}\right]  .
\end{array}
\label{elirozn}%
\end{equation}
Now, let us replace the variables $q_{N+1},\ldots,q_{n}$ in the first $N$
components of the \emph{last} $s$ systems in (\ref{Kilq}). That leads to $s$
highly nonlinear (purely differential) evolutionary equations of the form%
\begin{equation}
\overline{q}_{t_{r}}=\overline{Z}_{r}^{N}\left[  \overline{q}\right]
\text{\ \ \ \ }r=n-s+1=N+1,\ldots n \label{lasts}%
\end{equation}
where as before $\overline{q}=\left(  q_{1},\ldots,q_{N}\right)  ^{T}$ but
with new, purely differential, vector fields $\overline{Z}_{r}^{N}$. These
fields constitute in fact the first $s$ fields of the local cHD hierarchy.
Contrary to the nonlocal case, however, the first field of the hierarchy
appears as the \emph{last} equation in (\ref{lasts}) i.e. $\overline{q}%
_{t_{n}}=\overline{Z}_{n}^{N}\left[  \overline{q}\right]  $, the second field
is $\overline{q}_{t_{n-1}}=\overline{Z}_{n-1}^{N}\left[  \overline{q}\right]
$ and so on so that the fields of the hierarchy originate in (\ref{lasts}) in
the reverse order. We will therefore introduce a new notation and denote
\begin{equation}
\tau_{p}=t_{n-p+1},\text{ and }\overline{X}_{p}^{N}=\overline{Z}_{n-p+1}%
^{N},\ \ \text{ }p=1,\ldots,n-1 \label{ns}%
\end{equation}
so that $\overline{q}_{t_{n}}=\overline{Z}_{n}^{N}\left[  \overline{q}\right]
$ reads $\overline{q}_{\tau_{1}}=\overline{X}_{1}^{N}\left[  \overline
{q}\right]  $ and so on. The sequence (\ref{lasts}) becomes therefore%
\begin{equation}
\overline{q}_{\tau_{r}}=\overline{X}_{r}^{N}\left[  \overline{q}\right]
,\text{ \ \ }r=1,\ldots,s\text{.} \label{przep}%
\end{equation}

A theorem analogous to Theorem \ref{incrs} explains that this procedure leads
to a hierarchy.

\begin{theorem}
The vector fields in (\ref{przep}) do not depend on $s$ in the sense that if
we increase $s$ to $s+1$ then the above elimination procedure produces the
same sequence (\ref{przep}) of evolutionary systems plus a new system
$\overline{q}_{\tau_{s+1}}=\overline{X}_{s+1}^{N}\left[  \overline{q}\right]
$ at the end of the sequence (\ref{przep}) (i.e. at the beginning of the
sequence (\ref{lasts})).
\end{theorem}

\begin{proof}
Consider the $s$ systems (\ref{lasts}) and increase $s$ to $s+1$ keeping $N$
constant. We have then to take $n+1$ instead of $n$ in our elimination
procedure. Since, according to Proposition \ref{nz2}, the functions
$\gamma_{i,1}^{\left(  N\right)  },\gamma_{i,2}^{\left(  N\right)  }%
,\gamma_{1}^{\left(  N\right)  }$ do not depend on $n$ the functions
$f_{i}^{(N)}$ do not depend on $n$ either. It means that increasing $n$ to
$n+1$ (and keeping $N$ constant) turns the equations (\ref{elirozn}) into%
\[%
\begin{array}
[c]{l}%
q_{n+1}=f_{1}^{(N)}\left[  q_{1},\ldots,q_{N}\right] \\
\vdots\\
q_{N+2}=f_{n}^{(N)}\left[  q_{1},\ldots,q_{N}\right] \\
q_{N+1}=f_{n+1}^{(N)}\left[  q_{1},\ldots,q_{N}\right]
\end{array}
\]
and at the same time the the structure of the last $s$ equations in
(\ref{Kilq}) changes so that $q_{n}$ is replaced by $q_{n+1}$, $q_{n-1}$ is
replaced by $q_{n}$ and so on until $q_{N+2}$. It means that the last $s$
equations in the (extended to $n+1$) sequence (\ref{lasts}) will after
elimination remain the same while a new equations originates - this time
before (with lowest $r$) the other $s$ ones.
\end{proof}

Thus, by taking appropriate $s$ we can produce on demand an arbitrary (finite)
number of evolutionary vector fields%
\begin{equation}
\overline{q}_{\tau_{r}}=\overline{X}_{r}^{N}\left[  \overline{q}\right]
,\text{ \ \ }r=1,2\ldots\label{celrozn}%
\end{equation}
and due to same argument as in the nonlocal case, these vector fields all
mutually commute:%
\[
\left[  \overline{X}_{i}^{N},\overline{X}_{j}^{N}\right]  =0\text{ for all
}i,j=1,2\ldots
\]
The described procedure leads in fact to multicomponent local Harry Dym hierarchy.

\begin{example}
Let us first produce the first $s=2$ flows of the standard Harry Dym hierarchy
i.e. with $N=1$. We have $n=s+N=3$. Consider the Lagrangian%
\[
L^{n,n-N,-n}=L^{3,2,-3}=\frac{1}{4}q_{1,x}^{2}-\frac{q_{2,x}q_{3,x}}{2q_{3}%
}+\frac{q_{3,x}^{2}q_{2}}{4q_{3}^{2}}+\frac{q_{1}}{4q_{3}^{2}}-\frac{q_{2}%
^{2}}{4q_{3}^{3}}%
\]
and the corresponding Euler-Lagrange equations (\ref{ELrozn}). They attain the
form%
\begin{align*}
E_{1}\left(  L^{3,2,-3}\right)   &  \equiv\frac{1}{4q_{3}^{2}}-\frac{1}%
{2}q_{1,xx}=0\\
E_{2}\left(  L^{3,2,-3}\right)   &  \equiv-\frac{q_{2}}{2q_{3}^{3}}%
-\frac{q_{3,x}^{2}}{4q_{3}^{2}}+\frac{q_{3,xx}}{2q_{3}}=0
\end{align*}
and can thus easily be solved with respect to $q_{2}$ and $q_{3}$ yielding
(\ref{elirozn}) in the explicit form%
\begin{align*}
q_{3}  &  =q_{3}[q_{1}]=\left(  2q_{1,xx}\right)  ^{-1/2}\\
q_{2}  &  =q_{2}[q_{1}]=\tfrac{1}{2}\left(  5q_{1,xxx}^{2}-4q_{1,xx}%
q_{1,xxxx}\right)  \left(  2q_{1,xx}\right)  ^{-7/2}%
\end{align*}
Substituting these expressions to the first (since $N=1$) component of the
last $s=2$ Killing systems of the sequence (\ref{Kilq}) we obtain the
following two commuting flows:%
\[
q_{1,t_{2}}=\left(  q_{2}[q_{1}]\right)  _{x}\text{ \ , \ \ }q_{1,t_{3}%
}=\left(  q_{3}[q_{1}]\right)  _{x}%
\]
or%
\[
q_{1,\tau_{1}}=\left(  q_{3}[q_{1}]\right)  _{x}=\overline{X}_{1}^{1}\text{
\ \ },\text{ \ \ }q_{1,\tau_{2}}=\left(  q_{2}[q_{1}]\right)  _{x}%
=\overline{X}_{2}^{1}%
\]
(with the differential functions $q_{2}[q_{1}]$ and $q_{3}[q_{1}]$ given as
above) i.e. the first two members of the well known local Harry Dym hierarchy.
\end{example}

\begin{example}
Let us now produce the first $s=2$ flows of the $N=2$-component Harry Dym
hierarchy, we need therefore $n=s+N=4$. The Euler-Lagrange equations
(\ref{ELrozn}) for the Lagrangian \thinspace$L^{n,n-N,-n}=L^{4,2,-4}$ attain
the form%
\begin{align*}
E_{1}\left(  L^{4,2,-4}\right)   &  \equiv\frac{1}{4q_{4}^{2}}+\frac{1}%
{2}q_{1}q_{1,xx}+\frac{1}{4}q_{1,x}^{2}-\frac{1}{2}q_{2,xx}\\
E_{2}\left(  L^{4,2,-4}\right)   &  \equiv-\frac{q_{3}}{2q_{4}^{3}}-\frac
{1}{2}q_{1,xx}%
\end{align*}
that is soluble with respect to $q_{3}$ and $q_{4}$ yielding%
\begin{align*}
q_{4}  &  =q_{4}[q_{1,}q_{2}]=-w^{-1/2}\equiv-\left(  2q_{2,xx}-q_{1,x}%
^{2}-2q_{1}q_{1,xx}\right)  ^{-1/2}\\
q_{3}  &  =q_{4}[q_{1,}q_{2}]=-q_{1,xx}w^{-3/2}%
\end{align*}
Substituting these functions to the first $N=2$ components of the last $s=2$
Killing systems of the sequence (\ref{Kilq}) (with $n=4$) yields the desired
flows%
\begin{equation}
\frac{d}{d\tau_{1}}\left(
\begin{array}
[c]{c}%
q_{1}\\
q_{2}%
\end{array}
\right)  =\overline{X}_{1}^{2}\equiv\left(
\begin{array}
[c]{c}%
\left(  w^{-1/2}\right)  _{x}\\
q_{1}\left(  w^{-1/2}\right)  _{x}-w^{-1/2}q_{1,x}%
\end{array}
\right)  \label{rozn1}%
\end{equation}
and%
\begin{equation}
\frac{d}{d\tau_{2}}\left(
\begin{array}
[c]{c}%
q_{1}\\
q_{2}%
\end{array}
\right)  =\overline{X}_{2}^{2}\equiv\left(
\begin{array}
[c]{c}%
\left(  q_{1,xx}w^{-3/2}\right)  _{x}\\
q_{1,xx}w^{-3/2}q_{1,x}-q_{1}\left(  q_{1,xx}w^{-3/2}\right)  _{x}+\left(
w^{-1/2}\right)  _{x}%
\end{array}
\right)  \label{rozn2}%
\end{equation}

\end{example}

Our parametrization of Harry Dym hierarchy differs from the parametrization
given in \cite{AF2}. Generally speaking, the hierarchy (\ref{celrozn}) is
transformed into the multicomponent Harry Dym hierarchy presented in
\cite{AF2} through a complex version of the map (\ref{mapka})%

\begin{equation}
u_{r}=-iE_{N-r+1}\left(  L^{N,0,0}\right)  \text{, \ \thinspace}%
r=1,\cdots,N\text{, \ \ \ }i^{2}=-1. \label{mapkacom}%
\end{equation}
For example, in the$\ u$-variables the system (\ref{rozn1}) attains the form%
\[
\frac{d}{d\tau_{1}}\left(
\begin{array}
[c]{c}%
u_{1}\\
u_{2}%
\end{array}
\right)  =\overline{X}_{1}^{2}\left[  u\right]  \equiv\left(
\begin{array}
[c]{c}%
\frac{1}{4}\left(  u_{2}^{-1/2}\right)  _{xxx}\\
u_{1}\left(  u_{2}^{-1/2}\right)  _{x}+\frac{1}{2}u_{2}^{-1/2}u_{1,x}%
\end{array}
\right)
\]
that is exactly the flow (24a) in \cite{AF2}.

We will now formulate a theorem corresponding to Theorem \ref{main}, i.e. we
will generate a wide class of solutions of the hierarchy (\ref{celrozn}).

\begin{theorem}
\label{mainrozn}For any $\beta\in\{0,\ldots,n-1\}$ the functions $\lambda
_{i}=\lambda_{i}(t_{1},\ldots,t_{n})$ given implicitly by%
\begin{equation}
t_{i}+c_{i}=\pm\frac{1}{2}\sum_{r=1}^{n}%
{\displaystyle\int}
\frac{\lambda_{r}^{n-i}}{\sqrt{\lambda_{r}^{-N+\beta}\left(
{\textstyle\sum\nolimits_{j=1}^{n}}
a_{j}\lambda_{r}^{n-j}+\frac{1}{4}\lambda_{r}^{-\beta}\right)  }}d\lambda
_{r}\text{, \ \ \ }i=1,\ldots,n\text{.} \label{solscHD}%
\end{equation}
are such that the corresponding functions \thinspace$q_{i}=q_{i}(x=t_{1}%
,t_{2},\ldots,t_{n})$, $i=1,\ldots,N$, given by (\ref{defq}) are solutions of
the first $n-\beta$ ($n-1$ for $\beta=0,1$) equations of the $N$-component
integral cHD hierarchy (\ref{celrozn}). The variables $t_{\beta+1}%
=\tau_{n-\beta},\ldots,t_{n}=\tau_{1}$ ($t_{2},\ldots,t_{n}$ for $\beta=0,1$)
are evolution parameters (dynamical times) while the remaining $t_{i}$'s are
free parameters.
\end{theorem}

We will not prove this theorem here as its proof resembles the proof of
Theorem \ref{main}. \ Comparing Theorems \ref{main} and \ref{mainrozn} we can
see that the solutions (\ref{solsintcHD}) and (\ref{solscHD}) are for
$\beta=1,\ldots,n-1$ related through the transformation $\beta\rightarrow
n-\beta$, $\varepsilon\rightarrow-\varepsilon$. i.e. every solution
(\ref{solsintcHD}) for $\beta=1,\ldots,n-1$ coincides, after changing
$\varepsilon\rightarrow-\varepsilon$, with the solution (\ref{solscHD}) with
$\beta^{\prime}=n-\beta$. It also means that the nonlocal flow $\overline
{q}_{t_{n-\beta+1}}=\overline{Z}_{n-\beta+1}^{N}\left[  \overline{q}\right]  $
and the local flow $\overline{q}_{\tau_{\beta}}=\overline{X}_{\beta}%
^{N}\left[  \overline{q}\right]  $ share the same family of solutions, namely
(\ref{solsintcHD}) (or (\ref{solscHD}) with $\beta^{\prime}=n-\beta$ and with
$\varepsilon^{\prime}=-\varepsilon$).

It turns out that (\ref{solscHD}) cannot be explicitly solved. However, by
taking all $a_{i}=0$ in (\ref{solscHD}) (which yields the so called
zero-energy solutions) we can obtain interesting implicit solutions to our
hierarchy (\ref{celrozn}).

\begin{example}
Consider the solutions (\ref{solscHD}) with $N=2$, $n=3$ and with all
$a_{i}=0$. They have the form%
\begin{equation}
t_{i}+c_{i}=\pm\sum_{r=1}^{3}%
{\displaystyle\int}
\lambda_{r}^{4-i}d\lambda_{r}\text{, \ \ \ }i=1,2,3\text{.} \label{exlocal}%
\end{equation}
(the same for all $\beta$ since $\beta$-terms cancel after inserting $a_{i}%
=0$) and according to Theorem \ref{mainrozn} they solve the first $n-1=2 $
flows of the $N=2$-component cHD hierarchy (\ref{celrozn}) i.e. both the flows
(\ref{rozn1}) and (\ref{rozn2}). Equations (\ref{exlocal}) after integrating
yield (remember that $t_{1}=x$; we also put all $c_{i}=0$ for simplicity of
the formulas)%
\begin{equation}
x=\frac{1}{4}%
{\textstyle\sum\nolimits_{i=1}^{3}}
\lambda_{i}^{4}\text{, \ \ }t_{2}=\frac{1}{3}%
{\textstyle\sum\nolimits_{i=1}^{3}}
\lambda_{i}^{3}\text{, \ \ }t_{3}=\frac{1}{2}%
{\textstyle\sum\nolimits_{i=1}^{3}}
\lambda_{i}^{2} \label{kaszana}%
\end{equation}
which can not be algebraically solved. However, similarly as in the nonlocal
case, we can embed (\ref{kaszana}) in the system%
\begin{align}
x  &  =\frac{1}{4}%
{\textstyle\sum\nolimits_{i=1}^{4}}
\lambda_{i}^{4}=\frac{1}{4}\left(  \rho_{1}^{4}-4\rho_{1}^{2}\rho_{2}%
+2\rho_{2}^{2}+4\rho_{1}\rho_{3}-4\rho_{4}\right) \nonumber\\
t_{2}  &  =\frac{1}{3}%
{\textstyle\sum\nolimits_{i=1}^{4}}
\lambda_{i}^{3}=\frac{1}{3}\left(  \rho_{1}^{3}-3\rho_{1}\rho_{2}+3\rho
_{3}\right) \label{kaszana2}\\
t_{3}  &  =\frac{1}{2}%
{\textstyle\sum\nolimits_{i=1}^{4}}
\lambda_{i}^{2}=\frac{1}{2}\left(  \rho_{1}^{2}-2\rho_{2}\right) \nonumber\\
t_{4}  &  =%
{\textstyle\sum\nolimits_{i=1}^{4}}
\lambda_{i}=\rho_{1}\nonumber
\end{align}
(where $\rho_{i}$ are symmetric polynomials in $\lambda_{i}$ so that
$q_{i}=\left(  -1\right)  ^{i}\rho_{i}$) in the sense that putting
$\lambda_{4}=0$ (so that $\rho_{4}=0$ since $\rho_{4}=\lambda_{1}\lambda
_{2}\lambda_{3}\lambda_{4}$; the righ hand sides of (\ref{kaszana2}) are again
due to (\ref{Newton})) in (\ref{kaszana2}) we obtain (\ref{kaszana}). The
equations (\ref{kaszana2}) can be explicitly solved yielding.%
\begin{align*}
q_{1}  &  =-\rho_{1}=-t_{4}\\
q_{2}  &  =\rho_{2}=-t_{3}+\frac{1}{2}t_{4}^{2}\\
q_{3}  &  =-\rho_{3}=-t_{2}-\frac{1}{6}t_{4}^{3}+t_{3}t_{4}\\
q_{4}  &  =\rho_{4}=-x+\frac{1}{24}t_{4}^{4}-\frac{1}{2}t_{4}^{2}t_{3}%
+\frac{1}{2}t_{3}^{2}+t_{2}t_{4}%
\end{align*}
Thus, the functions $q_{i}(x,t_{2},t_{3})$ given implicitly by%
\begin{align*}
q_{1}  &  =-\rho_{1}=-t_{4}(x,t_{1},t_{2})\\
q_{2}  &  =\rho_{2}=-t_{3}(x,t_{1},t_{2})+\frac{1}{2}t_{4}(x,t_{1},t_{2})^{2}%
\end{align*}
where $t_{3}(x,t_{1},t_{2})$ and $t_{4}(x,t_{1},t_{2})$ are any pair of
functions identically satisfying the condition
\[
0=-x+\frac{1}{24}t_{4}^{4}-\frac{1}{2}t_{4}^{2}t_{3}+\frac{1}{2}t_{3}%
^{2}+t_{2}t_{4}%
\]
solve both (\ref{rozn1}) and (\ref{rozn2}).
\end{example}

\section{Conclusions}

In this article we presented a novel method of obtaining multicomponent Harry
Dym hierarchy (both its local and nonlocal part) as well as wide classes of
its solutions, from a family of finite dimensional separable systems
(St\"{a}ckel systems of Benenti type). This method has been previously applied
to coupled Korteveg-de Vries hierarchy where it produced novel rational
solutions and also a family of implicit solutions. In the case of cHD
hierarchy discussed here, the method produces among others rational and
implicit solutions in case of nonlocal hierarchy and implicit solutions of the
local part. In addition, the method produces wide families of other solutions
that are to be exploited elsewhere. It also indicates the existence of common
solutions of local and nonlocal cHD systems.

Our method can hopefully be extended to other systems, for example by taking
more general separation relations than relations (\ref{BenSR}).

\section{Appendix}

We prove here Theorem \ref{main}. We start with the case $\beta=0$. For
$\beta=0$ the solutions (\ref{solsintcHD}) are just solutions (\ref{gensol})
with our choice of $m$ and $k$, namely $m=-N,k=0$. The functions
\begin{equation}
q_{1}(x=t_{1},t_{2},\ldots,t_{n}),\ldots,q_{n}(x=t_{1},t_{2},\ldots,t_{n})
\label{00}%
\end{equation}
obtained from (\ref{solsintcHD}) (with $\beta=0$) through (\ref{defq}) satisfy
thus all $n-1$ Killing systems (\ref{Kilq}). Moreover they satisfy all the
equations (\ref{EL1}) and thus also all the equations (\ref{eli2}) used in our
elimination procedure. This means that we are free to use any part of
(\ref{EL1}) or (\ref{eli2}) to perform an elimination of variables in
(\ref{Kilq}). Such elimination thus leads to new equations that are satisfied
by those functions from the set (\ref{00}) that survive the elimination. Now,
we know that replacing the variables $q_{N+1},\ldots,q_{n} $ in the first $N$
components of the first $s-1=n-N$ equations (\ref{Kilq}) by the functions
given by (\ref{eli2}) leads to the first $s-1$ flows of the hierarchy
(\ref{cel1}). That means precisely that the first $N$ functions in (\ref{00})
\begin{equation}
q_{1}(x=t_{1},t_{2},\ldots,t_{n}),\ldots,q_{N}(x=t_{1},t_{2},\ldots,t_{n})
\label{aa}%
\end{equation}
satisfy the first $s-1=n-N$ equations in (\ref{cel1}). We will now show that
they actually solve the first $n-1$ equation in (\ref{cel1}). Consider the
next flow $\overline{q}_{t_{s+1}}=\overline{Z}_{s+1}^{N}\left[  q\right]  $ in
(\ref{cel1}). In order to obtain this flow, we have to perform the elimination
of variables $q_{N+1},\ldots,q_{n},q_{n+1}$ in the flow $q_{t_{s+1}}%
=Z_{s+1}^{n+1}\left[  q\right]  $ through (\ref{eli2}) written for $n+1$
instead of $n$ i.e. obtained from solving (\ref{EL2}). This elimination is
therefore performed with the help of the same functions $q_{i}=q_{i}%
[q_{1,}\ldots,q_{N}]$ as for $n$ plus a new function $q_{n+1}=f_{n-N+2}\left[
q_{1},\ldots,q_{N}\right]  $. However, $\left(  Z_{s+1}^{n+1}\left[  q\right]
\right)  ^{j}=\left(  Z_{s+1}^{n}\left[  q\right]  \right)  ^{j}$ for all
$j=1,\ldots,N-1$ (it follows from (\ref{Kilq})) while $\left(  Z_{s+1}%
^{n+1}\left[  q\right]  \right)  ^{N}$ contains the additional variable
$q_{n+1\,}$not present in $\left(  Z_{s+1}^{n}\left[  q\right]  \right)  ^{N}%
$. It means that solutions (\ref{aa}) will certainly satisfy the first $N-1$
components in $\overline{q}_{t_{s+1}}=\overline{Z}_{s+1}^{N}\left[  q\right]
$. Further, since $E_{N+1}\left(  L^{n+1,-N,0}\right)  =-\frac{1}{2}%
q_{n+1,xx}+E_{N}\left(  L^{n,-N,0}\right)  $, the function $q_{n+1}%
=f_{n-N+2}\left[  q_{1},\ldots,q_{N}\right]  $ is (after choosing both
integration constants equal to zero) identically equal to zero on the
solutions (\ref{aa}). That means that on the solutions (\ref{aa}) we have
$\left(  Z_{s+1}^{n+1}\left[  q\right]  \right)  ^{N}=\left(  Z_{s+1}%
^{n}\left[  q\right]  \right)  ^{N}$ which means indeed that (\ref{aa}) solves
$\overline{q}_{t_{s+1}}=\overline{Z}_{s+1}^{N}\left[  q\right]  $. By
expanding this argument, the functions $q_{n+1},q_{n+2},\ldots,q_{n+N-1}$
obtained from (\ref{EL1}) with $n$ replaced by $n^{\prime}=n+N-1$ i.e. from
the $n^{\prime}-N=n-1$ equations
\begin{equation}%
\begin{array}
[c]{l}%
E_{N+1}\left(  L^{n^{\prime},-N,0}\right)  \equiv-\frac{1}{2}q_{n+N-1,xx}%
+\varphi_{n-1}[q_{1},...,q_{n+N-2}]=0,\\
E_{N+2}\left(  L^{n^{\prime},-N,0}\right)  \equiv-\frac{1}{2}q_{n+N-2,xx}%
+\varphi_{n-2}[q_{1},...,q_{n+N-3}]=0,\\
\vdots\\
E_{n^{\prime}-1}\left(  L^{n^{\prime},-N,0}\right)  \equiv-\frac{1}%
{2}q_{N+2,xx}+\varphi_{2}[q_{1},...,q_{N+1}]=0\\
E_{n^{\prime}}\left(  L^{n^{\prime},-N,0}\right)  \equiv-\frac{1}{2}%
q_{N+1,xx}+\varphi_{1}[q_{1},...,q_{N}]=0.
\end{array}
\label{EL3}%
\end{equation}
(which are necessary to obtain the first $n-1$ flows of (\ref{cel1})) are
identically zero on the solutions (\ref{aa}) which leads to the conclusion
that (\ref{aa}) indeed solve the first $n-1$ equations of (\ref{cel1}).

Assume finally that $0<\beta\leq n-1$. The functions (\ref{solsintcHD}) are
then the complete solution (as usual, through the map (\ref{defq})) of all the
Euler-Lagrange equations $E_{i}(L^{n,\beta-N,-\beta})$ associated with the
Lagrangian $L^{n,\beta-N,-\beta}$. As such, they still must solve all the
Killing systems (\ref{Kilq}). However, since $E_{i}(L^{n,\beta-N,-\beta
})=E_{i+\beta}(L^{n,-N,0})$ for $i=1,\ldots,n-\beta$ due to (\ref{wtyl}), for
any $\beta>1$ we lose the first $\beta-1$ equations in (\ref{EL3}) which means
that our proof works only for the first $n-\beta$ flows in (\ref{cel1}) - we
simply can not "blow up" $n$ to $n^{\prime}=n+N-1$ but only to $n^{\prime
\prime}=n+N-1-\beta$.

\section{Acknowledgement}

Both authors were partially supported by Swedish Research Council grant no VR
2009-414 and by Ministry of Science and Higher Education (MNiSW) of the
Republic of Poland research grant No.~N~N202~4049~33.

\end{document}